\documentclass[twocolumn]{revtex4} 
\usepackage{latexsym}
\usepackage{graphicx}
\usepackage{fancyhdr}
\renewcommand{\bar}[1]{\overline{#1}}

\begin{document}


\title{
Superfluidity in many fermion systems: Exact renormalisation group treatment}

\author{Boris Krippa}

\affiliation{School of Physics and Astronomy,
The University of Manchester, M13 9PL, Manchester, UK}

\begin{abstract}
The application of the exact renormalisation group to  
 symmetric as well as asymmetric many-fermion systems with a short-range attractive force is studied.
 Assuming an  ansatz for 
the effective action with effective bosons, describing pairing effects and 
  a set of approximate flow equations for 
the effective coupling including boson and fermionic fluctuations has been derived. 
 The  phase transition to a 
phase with broken symmetry is found at a critical value of the running 
scale.  The mean-field results are recovered  if   boson-loop effects are omitted.
The calculations with two different forms of the regulator are shown to lead to a 
 similar results. We find that, being quite small in  the case of the symmetric  
many-fermion system the corrections to  mean field approximation becomes more important with increasing
mass asymmetry.

\end{abstract}

\maketitle
\thispagestyle{fancy}

\section{Introduction}
There is a growing interest in applying the Exact Renormalisation Group (ERG) formalism 
to few- and many-body systems \cite{Bir,Kr1,Kr2} when the underlying interaction is essentially non-perturbative.
Regardless of the details all ERG-based approaches share the same distinctive feature,
 a successive elimination/suppression of some modes, resulting in effective interaction between the remaining degrees of freedom.
One specific way of implementing such a procedure is to eliminate modes by 
 applying a momentum-space blocking transformation
with some physically motivated cutoff. The effect of varying a cutoff is described by  nonlinear 
ERG evolution equations, which include the effect of the eliminated modes.In the following we will use the variant
of ERG based on the concept of Average Effective Action (AEA) \cite{Wet}.
 The corresponding evolution equation can written in the following general form

\begin{equation}
\partial_k\Gamma=-\frac{i}{2}\, Tr \left[(\partial_kR)\,
(\Gamma^{(2)}-R)^{-1}\right].
\end{equation}
Here $\Gamma^{(2)}$ is the second functional derivative of the AEA  taken with respect to all types of field included in the action.
and $R$ is a regulator which should suppress the contributions of states with momenta
less than or of the order of running scale $k$. To recover the full effective action we require $R(k)$ to vanish as $k\rightarrow$ 0, in other
respects its form is rather arbitrary. The concrete functional form of the regulator has no effect on physical results provided
 no approximations/truncations were made.By solving the ERG equations one can find a scale dependence of the coupling constants
 and thus determine the ERG flow.
 The whole approach is nonperturbative so that some physically motivated assumption about the functional form
of the effective action should be made
It is well known that the many-fermion system with attractive interaction favours the formation of the correlated fermion pairs
leading to the symmetry breaking and related phenomena.
Since we expect the appearance of the correlated fermion pairs in a physical 
ground state, we need to parametrise our effective action in a way that
can describe the qualitative change in the physics when this occurs.
A natural way to do this is to introduce a boson field whose vacuum 
expectation value (VEV) describes this correlated pair  \cite{Wein94} and study the evolution of this effective degrees of freedom. At  
the start of the RG evolution, the boson 
field is not dynamical and is introduced through a 
Hubbard-Stratonovich transformation of the four-point interaction.
As we integrate out more and more of the fermion degrees of freedom by 
running the cutoff scale $k$ to lower values, we generate dynamical terms in the bosonic
effective action.
 In this paper we treat both symmetric and asymmetric many fermion systems. The corresponding ansatz 
for the  boson-fermion effective action consists of the kinetic terms for boson and fermions and the interaction term and for two types 
of fermions can be written as
\begin{equation}
\Gamma[\mu,k] = \int d^4x\left(\Gamma_{B}[\mu,k]+\Gamma_{F}[\mu,k]+
\Gamma_{I}[k]\right)
\end{equation}
Here $\Gamma_{B(F)}$ is the boson (fermion)  part of AEA
\begin{equation}
\Gamma_{B} = \phi^\dagger\left(Z_\phi\, (i \partial_t +\mu_a + \mu_b)
+\frac{Z_m}{2m}\,\nabla^2\right)\phi-U(\phi,\phi^\dagger)
\end{equation}

\begin{equation}
\Gamma_{F}= \sum_{i=a}^{b}\psi_{i}^\dagger\left( Z_{\psi,i} (i \partial_t+\mu_i)
+\frac{Z_{M,i}}{2M_i}\,\nabla^2\right)\psi_i
\end{equation}
and  $\Gamma_{I}$ is the interaction term
\begin{equation}
\Gamma_{I} = - Z_{g} \left(\frac{i}{2}\,\psi_{b}^{\rm T}\sigma_2\psi_{a}\phi^\dagger
-\frac{i}{2}\,\psi_{a}^\dagger\sigma_2\psi_{b}^{\dagger{\rm T}}\phi\right)
\end{equation}
M is the reduced mass of the fermion in vacuum and the factor $1/2m$ with $m=M_a + M_b$ in the 
boson kinetic term is chosen simply to make $Z_m$ dimensionless. The
coupling $Z_g$,  the 
wave-function renormalisations factors $Z_{\phi,\psi}$ and the kinetic-mass 
renormalisations factors $Z_{m,M}$ all run with on $k$, the scale of the regulator. Having in mind the future applications to the 
 crossover from BCS to BEC (where chemical potential becomes negative) we also let the chemical potentials $\mu_a$ and $\mu_b$ run,
 thus keeping the corresponding densities (and Fermi momenta $p_{F,i}$) constant.  
The bosons are , in principle, coupled to the chemical potentials via a quadratic term in $\phi$, 
but this can be absorbed into the potential by defining $\bar U=U-(\mu_1 +\mu_2) Z_\phi\phi^\dagger\phi$. 
The evolution equations include running of chemical potentials, effective potential and all couplings ($Z_\phi, Z_{m}, Z_{M,i}, Z_{\psi,i}, Z_g$).
 However, in this paper we allow to run only $Z_\phi$, parameters in the effective potential ($u's$ and $\rho_0$) and 
chemical potentials since this is the minimal set needed to include the effective boson dynamics.
The system with one type of fermion corresponds to the limit $M_a = M_b$.

We expand the effective potential about its minimum, $\phi^\dagger\phi=\rho_0$, so that the coefficients $u_i$ are defined at $\rho=\rho_0$,  
\begin{equation}
\bar U(\rho)= u_0+ u_1(\rho-\rho_0)
+\frac{1}{2}\, u_2(\rho-\rho_0)^2
+\frac{1}{6}\, u_3(\rho-\rho_0)^3+\cdots,
\label{eq:potexp}
\end{equation}
where we have introduced $\rho=\phi^\dagger\phi$. Similar expansion can be written for the renormalisation factors.
 The coefficients of the expansion run with the scale.  The phase of the system is determined by the
 coefficient $u_1$.
We start evolution at high scale where the system is in the symmetric phase so that  $u_1 > 0$. When the running scale  becomes 
comparable with the pairing scale (close to  average Fermi-momentum) the system undergoes the phase transition to the phase with broken symmetry,
energy gap etc. The point of the transition corresponds to the scale where   $u_1 = 0$. The bosonic excitations in the gapped phase are 
gap-less Goldstone bosons. Note, that  in this phase the minimum of the potential will also run with the scale $k$ so that the value  $\rho_0 (k \rightarrow 0)$
determines the physical gap.

The important part of any ERG treatment is the choice of the regulator. Ideally, the physical results should not depend on this choice. However, 
some sort of truncations and approximations should always be made in  real calculations  to render the system of the resulting 
evolution equations solvable so that the convenient choice of the regulator is the question of significant practical importance.
In our approach the boson regulator  has the structure 
\begin{equation}
{\bf R}_B = R_B diag(1,1),  
\end{equation}
and the fermion regulator for both types of fermions 
has the structure 
\begin{equation}
{\bf R}_{F,i} = sgn(\epsilon_{i}(q) - \mu_i) R_{F,i}(q,\mu_i,k) diag(1,-1)
\end{equation}
 Note that this regulator is positive for particle states above the Fermi surface and negative for the hole states below the Fermi surface.
The function $R_F$ should suppress the contributions 
of states with momenta near the Fermi surface, $|q-p_F|\sim k$. Once a 
a large gap has appeared in the fermion spectrum, there are no low-energy fermion
excitations and so the fermionic regulator plays little further role. However, 
while the gap is zero or small, it is crucial that the sign of the regulator
match that of the energy, $q^2/2M - \mu$, and hence it is $\mu$ which
appears in the sign functions.

The other important part of the ERG approach is fixing the boundary conditions which define the form of the AEA at some initial scale so that  
at some large starting scale $k=K$ we demand that the Lagrangian be equivalent to
a purely fermionic theory with the contact interaction
\begin{equation}
{\cal L}_i=-\frac{1}{4}\,C_0\left(\psi^\dagger\sigma_2\psi^{\dagger{\rm T}}\right)
\left(\psi^{\rm T}\sigma_2\psi\right).
\end{equation}
Here $C_0(K)$ is the strength of the energy-independent term in the effective NN
interaction in vacuum. This is evaluated 
at the scale $K$, using the same regularisation procedure as we apply in matter.
This equation  implies that $u_1$ and $g$ at the this scale are related by
\begin{equation}
C_0(K)=-\,\frac{ g(K)^2}{ u_1(K)}.
\end{equation}
Using this boundary condition we can relate the pairing phenomena in a many body invironment with the fermion-fermion interaction in vacuum.
 
The boson potential $\bar U$ is obtained by evaluating the effective action for
uniform boson fields. It evolves according to
\begin{equation}
\partial_k \bar U=-\frac{1}{{\cal V}_4}\,\partial_k\Gamma,
\end{equation}
where ${\cal V}_4$ is the volume of spacetime.
Substituting our expansion of $\bar U$, Eq.~(\ref{eq:potexp}), on the left-hand
side leads to a set of ordinary differential equations for the $u_n$.

For the boson wave-function renormalisation factor, $Z_\phi$, we need to consider
a time-dependent background field. Taking
\begin{equation}
\phi(x)=\phi_0+\eta e^{-ip_0t},
\label{eq:tdfield}
\end{equation}
where $\eta$ is a constant, we can get the evolution of $Z_\phi$ from
\begin{equation}
\partial_k Z_\phi=\frac{1}{{\cal V}_4}\left.\frac{\partial}{\partial p_0}
\left(\frac{\partial^2}{\partial\eta\partial\eta^\dagger}\,\partial_k\Gamma
\right)_{\eta=0}\,\right|_{p_0=0}.
\end{equation}
Let us first consider the results of the calculations in the case of symmetric  fermion matter.
We solve the evolution equations numerically with two types of cutoffs.
First, we use the smoothed step-function type of regulator (called hereafter as $R_{1F}$)
\begin{equation}
R_{1F}=\frac{k^2}{2M}\theta_{1}(q - p_{F},k,\sigma); R_{1B}=\frac{k^2}{2m}\theta_{1}(q,k,\sigma)
\end{equation}
where
\begin{equation}
\theta_{1}(q,k,\sigma)=\frac{1}{2 erf(1/\sigma)}\left[erf(\frac{q +k}{k \sigma})
 + erf(\frac{q - k}{k \sigma})\right]
\end{equation}
with $\sigma$ being a parameter determining the sharpness of the step.
Second, we use the sharp cutoff  function $R_{2F}$ which is  somewhat similar to one suggested
in Ref.\cite{Litim} for the pure  boson case and chosen  in a rather peculiar way
 to make the calculations as simple as possible
\begin{eqnarray}
R_{2F}=\frac{k^2}{2M}[((k + p_\mu)^{2}- q^2)\theta(p_{\mu}+k -q) +\nonumber\\
 ((k + p_\mu)^{2}+ q^2 -2p^{2}_\mu)\theta(q - p_{\mu}+k)],
\end{eqnarray}
\begin{equation}
R_{2B}=\frac{k^2}{2m}(k^2 - q^2)\theta(k -q), 
\end{equation}
where $p_{\mu} = (2 M\mu)^{1/2}$.
The fermion sharp cutoff consists of two terms which result in modification of the particle  and hole propagators
 respectively. 
The hole term is further modified to suppress the contribution from the surface terms, which may bring in the dangerous dependence
 of the 
regulator on the cutoff scale even at the vanishingly small $k$. As an example, we focus on the parameters relevant to neutron matter:
$M=4.76 fm^{-1}, p_{F}=1.37 fm^{-1}$. Let's first discuss the results obtained with the smooth cutoff $R_1$. We found that the value of the
physical gap  is  practically independent of either the values of the width parameter $\sigma$ (varied within some range) or
the starting scale $K$ provided $K > 5 fm^{-1}$. The results of the calculations are shown on Fig. 1 
\begin{figure}
\includegraphics[width=9cm,  keepaspectratio,clip]{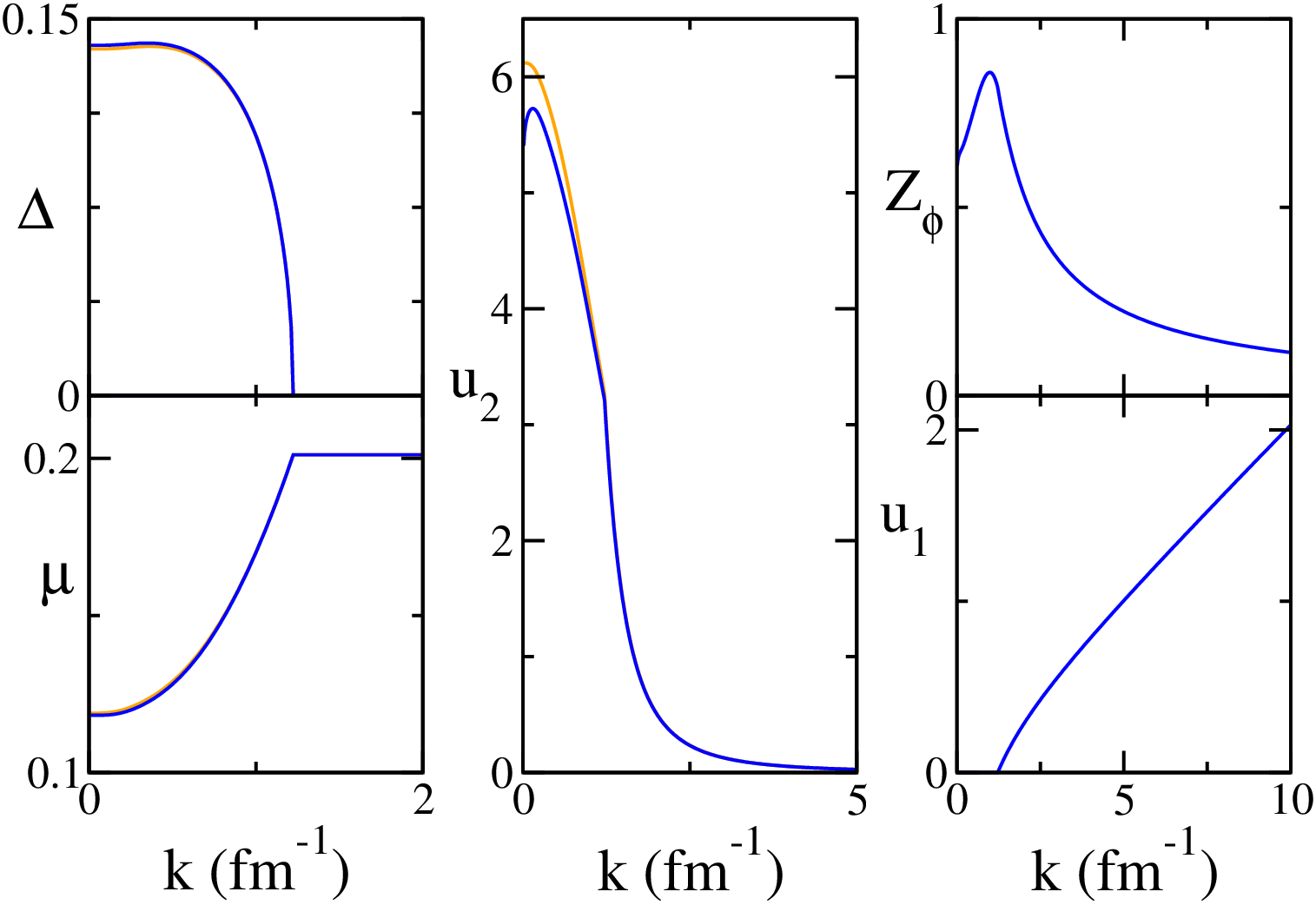}
\caption{\label{fig:running}Numerical solutions to the evolution equations
for infinite $a_0$ and $p_F=1.37$~fm, starting from $K=16$~fm$^{-1}$.
We show the evolution of all relevant 
parameters for the cases of fermion loops only (orange/grey lines), and  
of bosonic loops with a running $Z_\phi$ (blue/black lines). All 
quantities are expressed in appropriate powers of fm$^{-1}$.}
\end{figure}
At starting scale
 the system is in the symmetric phase and remains in this phase until $u_1$ hits zero at $k_{crit}\simeq 1.2 fm^{-1}$ where the artificial second
order phase transition to a broken phase occurs and the energy gap is formed. Already at $k \simeq 0.5$ the running scale has essentually no effect on
 the gap. 
 It is worth mentioning that we found very small (on the level of 1$\%$)
 contribution to the gap from the boson loops, due to cancellations between the direct contributions to the running of the gap and indirect ones via $u_2$.
 The boson loops play much more important role  in  the evolution of $u_2$ and $Z_\phi$. In fact, they drive
both couplings to zero at $k\rightarrow 0$. We note however, that the effect of the boson loops for the gap may still  be more visible if the evolution of
 the other couplings is included.\\
 The results obtained with sharp cutoff regulator which are shown on Fig.2.
\begin{figure}
\begin{center}
\includegraphics[width=9cm,  keepaspectratio,clip]{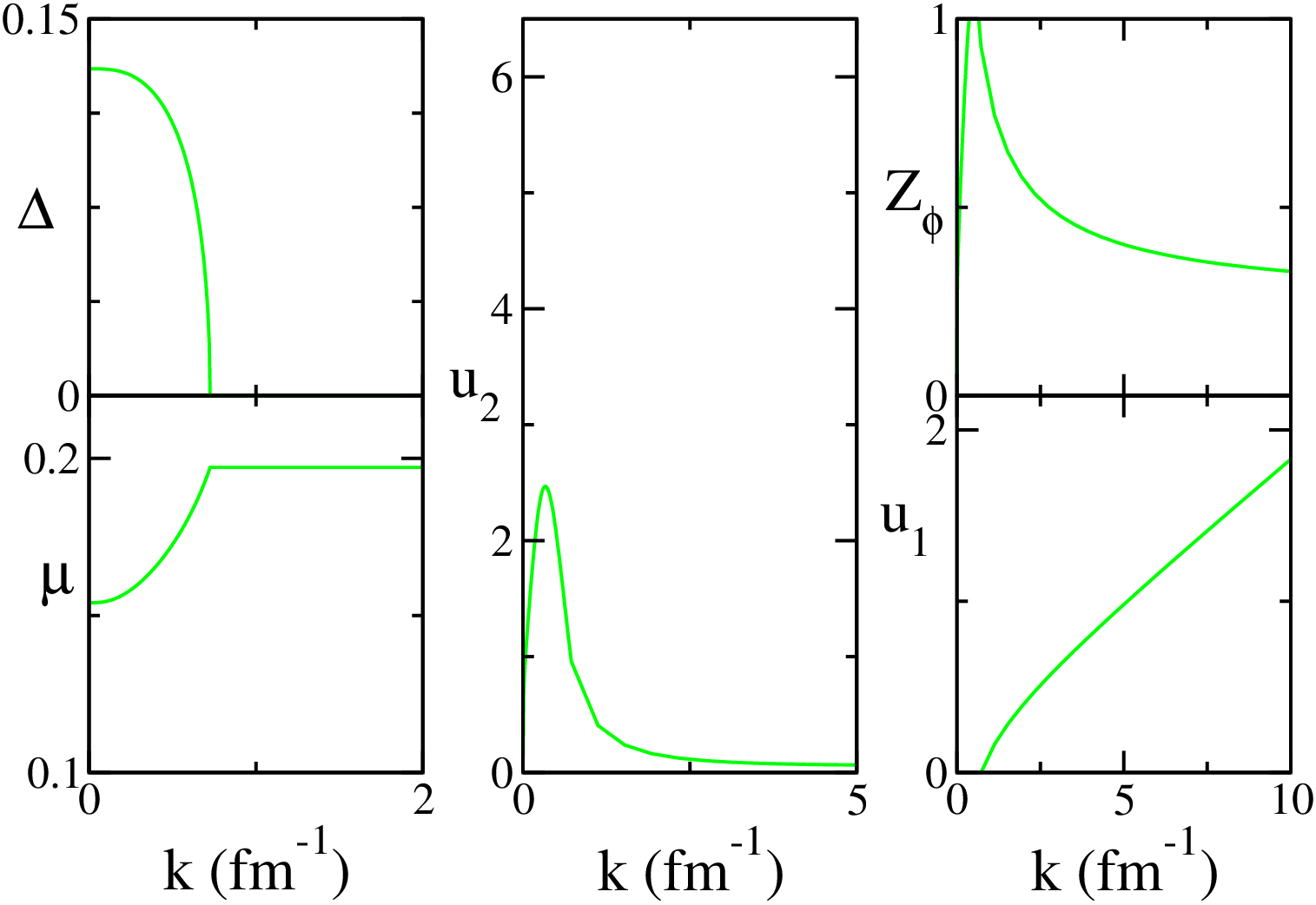}
\end{center}
\caption{\label{fig:running2}Evolution of the parameters when the sharp cut-off is used}
\end{figure}
 One immediate observation is that that the results become starting scale independent
as long as $K > 5 fm^{-1}$ similar to the case with the smooth cutoff. However, the artificial phase transition occurs at lower values of the running scale
$k \simeq 0.7 fm^{-1}$. At approximately $k \simeq 0.2 fm^{-1}$ the value of the gap becomes scale independent. One notes that the
 curves obtained with different regulators and describing
the evolution of the gap, being rather different at intermediate scales, approach each other with decreasing scale 
resulting in very close values for the physical gap. 
This is an encouraging results taking into account that, although the hypothetical exact results must be independent of the choice of the regulator in practice
 it is not garanteed at  all given the  assumed ansatz  for the effective action and truncations  made.
 The same conclusion also holds for the other quantities. The couplings $Z_{\phi}$ and $u_2$ first grow with scale and 
then start decreasing eventually coming to zero. Chemical potential begins to decrease at the point of phase transition and becomes scale independent at
$k\simeq 0.2 fm^{-1}$. However, in this case the numerical values of the chemical potentials obtained with different regulators differ
 by approximately $20\%$ so that this quantity is more sensitive to the details of effective action and to the trancations made.

Let us now discuss the results for the   many fermion system with two fermion species.  For simplicity we consider the case of
 the hypothetical ``nuclear'' matter with short range attractive interaction 
between  two types of fermions, light and heavy,
and study the behaviour of the energy gap as the function of the mass asymmetry. We choose the Fermi momentum to be  $p_{F}=1.37 fm^{-1}$.
 One notes that the formalism is applicable to any type of a many-body system with two fermion species
 from quark matter to fermionic atoms so that the    hypothetical asymmetrical ``nuclear'' matter is simply chosen as  a study case.
We assume that $M_a < M_b$, where $M_a$ is always the mass of the physical nucleon.

First we consider the case of the unitary limit with  the infinite scattering length. 
The results of our calculations for the gap 
 are shown on Fig. 3. 

\begin{figure}
\begin{center}
\includegraphics[width=8cm,  keepaspectratio,clip]{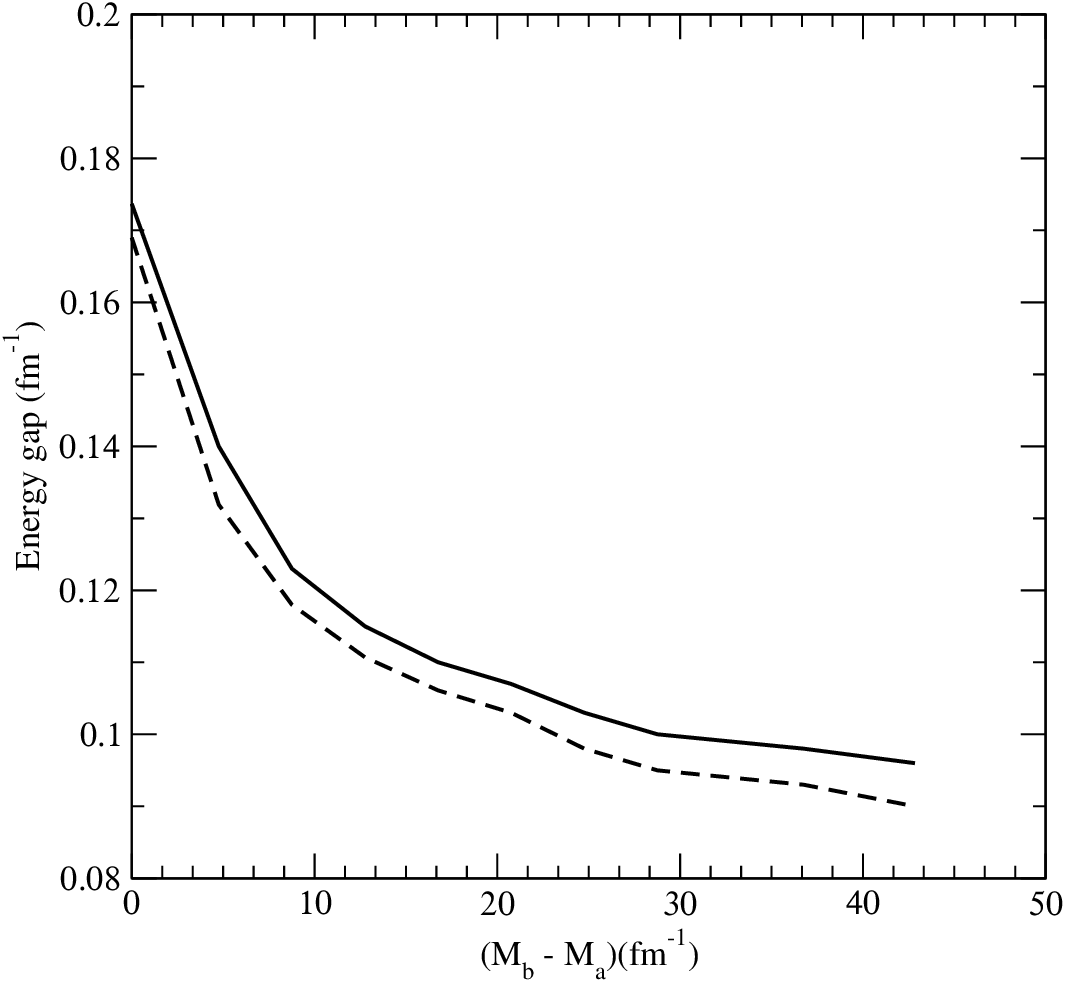}
\end{center}
\caption{\label{fig:running1}Evolution of the gap in the MF approach (dashed curve) and with boson loops (solid curve) in the unitary regime
$a = -\infty$ as a function of a mass asymmetry.}
\end{figure}

 We see from this figure  that  increasing  mass asymmetry leads to a decreasing gap that seems to be a natural result.
However, the effect of the boson loops is found to be   small. We found essentially no effect in symmetric phase, $2-4 \%$ corrections for the value of the 
gap in the broken phase and even smaller corrections for the chemical potential so that one can conclude that
the MF approach indeed provides the reliable description in the unitary limit for both small and large mass asymmetries.
It is worth mentioning that, similar to above considered case of the fermion matter with one type of fermion,  the
 boson contributions are more important for the evolution of $u_2$ where they drive  $u_2$ to zero as $k\rightarrow 0$ making the effective potential
convex in agreement with the general expectations. This tendency  retains in the unitary regime regardless of the mass asymmetry. 

We have also considered
the behaviour of the gap as the function of the parameter $p_F a$ for the cases of the zero asymmetry $M_a = M_b$ and the  maximal asymmetry $M_b = 10 M_a$.
The results are shown on Fig.4.
\begin{figure}
\begin{center}
\includegraphics[width=8.65cm,  keepaspectratio,clip]{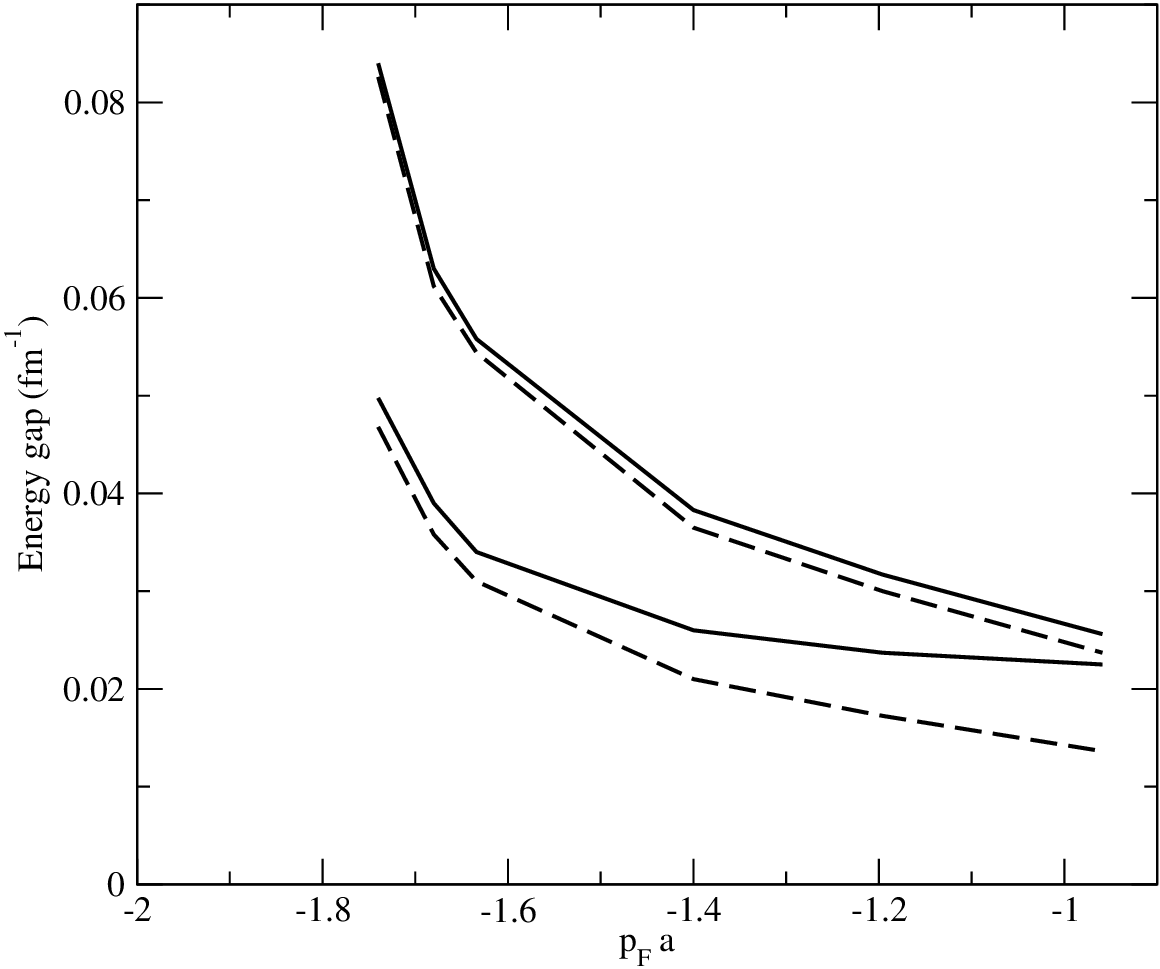}
\end{center}
\caption{\label{fig:running2}Evolution of the gap as a function of the parameter $p_F a$. The upper 
pair of the  curves corresponds to the calculations with no asymmetry in the MF approach (dashed curve) and with boson loops (solid curve) and the lower pair 
of the curves describes the results of calculations with the  maximal asymmetry when $M_b = 10 M_a$}.
\end{figure}

  One can see from  Fig.4 that in the  case of zero (or small) asymmetry  the corrections stemming from  boson loops
are small at all values of   the parameter $p_F a$ considered here (down to $p_F a = 0.94)$. On the contrary, when  $M_b = 10 M_a$ these corrections, being 
rather small at $p_F a \geq  2$ becomes significant ($\sim  40\%$) when the value of $p_F a$ decreases down to $p_F a \sim  1$. We found that  
at $p_F a \sim  1$ the effect of boson fluctuations becomes  $\sim  10\%$  already for $M_b = 5 M_a$. One can therefore conclude 
 that the regime of large mass asymmetries, which starts  approximately at  $M_b > 5 M_a$, 
 moderate scattering length and/or the Fermi momenta is the one where the MF description becomes less accurate so that the calculations going beyond the MFA
are needed. One might expect that the deviation from the mean field results could even be stronger in a general case of a large mass asymmetry and 
the mismatched Fermi surfaces but the detailed conclusion can only be drawn after the actual calculations are performed.

 We were not able to follow the evolution of the system at
 small gap (or small $p_F a$) because of the non-analyticity of the effective action
 in this case. This non-analyticity of the effective action can explicitly be demonstrated in the mean-field approximation. The flow equations
can be solved analytically in this case and one can see from the solution, which has a closed-form expression in terms of an associated
Legendre function, $P_l^m(y)$ at $k=0$,
  that 
 the fermion loops contain a term
$\phi^\dagger\phi\log(\phi^\dagger\phi)$. It remains to be seen whether, within the given ansatz, the full solution of the system of the 
partial differential equations for the effective potential and running couplings is required to trace the 
evolution of the system in the case of small gaps. 

In summary, we have studied the pairing effect for the asymmetric fermion matter 
with two fermion species as a function of fermion mass asymmetry.
 We found that  regardless of the size of the fermion mass asymmetry
the boson loop corrections are small at large enough values  of $p_F a$ so that  the MFA provides a consistent description of the pairing effect 
in this case. However, when $p_F a \sim  1$ these corrections become significant at large asymmetries ($M_b > 5 M_a$) making the MFA inadequate. 
In this case it seems to be necessary to go beyond the mean field description.

There are several ways where this approach can further be developed. In the case of asymmetric systems 
the next natural step  would be to consider the case of the mismatched Fermi surfaces
 taking into account the possibility of formation of Sarma \cite {Wil} , mixed \cite {Bed, Cal} and/or LOFF \cite {Lar}
phases, exploring the importance of the boson loop for the 
stability of those phases
and applying the approach to the real physical systems, for example fermionic atoms.
Work in this direction is in progress. The other important extension of this approach would be to take into account running of all
couplings of the average effective action, use different type of cut-off function, preferably the smooth one and inlude  both  particle-hole
channel and long range forces. The three body force effects \cite {Kr4},
 when the correlated pair interact
with the unpaired fermion may also be important, especially for non-dilute systems.
\bigskip
\begin{center}
{\bf Acknowledgments}
\end{center}
The author would like to thank Mike Birse, Niels Walet and Judith McGovern for very useful discussions. 
\bigskip
\bigskip

\end{document}